**Superconducting Condensation Energy of the Two-Dimensional Hubbard Model in the Large-Negative-$t'$ Region**

Kunihiko Yamaji[1*], Takashi Yanagisawa[1], Mitake Miyazaki[2], and Ryosuke Kadono[3]

[1] Electronics and Photonics Research Institute, AIST, Central 2, 1-1-1 Umezono, Tsukuba 305-8568, Japan

[2] Hakodate National College of Technology, 14-1 Tokura-cho, Hakodate 042-8501, Japan

[3] Institute of Materials Structure Science, KEK, Tsukuba 305-0801, Japan

**Abstract**

We compute the superconducting condensation energies $E_{cond}$ of Hg1201 and Tl2201 cuprates by applying the variational Monte Carlo method to the two-dimensional Hubbard model, first, with the specific band parameters $t' = -2t'' = -0.25t$ appropriate for these highest-$T_c$ single-CuO$_2$-layer cuprates; $t$, $t'$ and $t''$ are the first-, second- and third-neighbor transfer energies, respectively. In the range of on-site Coulomb energy $U$ of $(7\sim10)t$ with optimal doping, we succeed in obtaining the bulk-limit $E_{cond}$ values by extrapolating the results. They sharply increase with increasing $U$ and become comparable to experimental values when $U \approx 9t$. Next, $E_{cond}$ values at $t' = -2t'' = -0.18t$ demonstrate that with a fixed $U$ the bulk-limit $E_{cond}$ quickly increases with a decrease in $|t'| = 2t''$ in the large-$|t'|$ region. Finally, we argue that the remoteness of the apex oxygen from the planar Cu in the two cuprates is considered to increase $U$ and bring about the experimental magnitude of $E_{cond}$. The present scheme explains the observed correlation between $T_c$ and $t'$ among cuprates.

*E-mail address: yamaji-kuni@aist.go.jp

The mechanism of high-$T_c$ superconductivity (SC) in cuprates remains an important basic problem. Many theories on this mechanism are based on the Coulomb interaction between electrons.[1-5] The simplest model of this type is the two-dimensional (2D) Hubbard model with a substantial on-site Coulomb energy $U$. Since the variational Monte Carlo (VMC) method is suitable for treating such correlated electron systems, we employed this method for studying the possible occurrence of SC in this model by computing the SC condensation energy $E_{cond}$. The magnitude of $U$ in reference to the nearest-neighbor transfer energy $t$ is a serious issue regarding the theory. Anderson gave a $U$ estimate of ~8 (hereafter, $t$ is the energy unit in this paper).[1] From the analysis of the experimental spin-wave dispersion, Coldea et al. derived $U \approx 7$ for non-doped $La_2CuO_4$.[6] In previous computational works[7-9], by comparing the results of SC $E_{cond}$ with experimental values, we tentatively chose $U = 6$. Our $U$ was more moderate in comparison with the other values out of consideration for the requirement that the SC parameter space for LSCO should survive against the spin-density-wave (SDW) phase.

Band parameters are also important parameters in this approach, which are represented by $t'$ and $t''$, second- and third-neighbor transfer energies, respectively, on the square lattice. Band theories and also experimental works gave their values distributed in a considerably wide range even within cuprates. They are known to take $t' = -0.05 \sim -0.30$ and $t'' \sim -t'/2$. In our previous work[9] we tried to determine the dependences of $E_{cond}$ on $t'$ and $t''$ in a wide and realistic range of $t'$ and $t''$ with a fixed relation $t'' = -t'/2$. The latter relation is justified as a simplified approximation in ref. 10. The doping level is fixed at an optimal value at approximately 16% hole doping throughout our work to avoid the complication associated with the pseudo-gap problem. Computations were carried out for large square lattices of the lattice sizes from 10×10 up to 24×24 and the bulk-limit SC $E_{cond}$ was determined by extrapolating $E_{cond}$ to an infinite lattice size. In the case of $t' = -2t'' = -0.05$ and also $-0.10$ with $U = 6$, we obtained in our previous work[9] bulk-limit values of SC $E_{cond}$ slightly larger than the typical experimental value 0.26 meV per active Cu site, or 0.00074 in units of $t = 0.35$eV, of YBCO.[11-13,8] We can say that the computed values are in fair agreement with the experimental ones in spite of the simplicity of the model and crudeness of the parameter values used. We consider that the $E_{cond}$ at $t' \approx -2t'' \approx -0.10$ for LSCO[14] gives the correct order of magnitude of the SC $E_{cond}$ of LSCO.

Hg1201 and Tl2201 have the highest $T_c$'s among single-$CuO_2$-layer cuprates: 98 and 93 K, respectively.[15] Experiments led to the estimation of $t' \approx -2t'' \approx -0.25$ for both compounds.[16,17] In our previous work with $U = 6$ employing a set of $t' = -2t'' \sim -0.20$, we found that the computed $E_{cond}$ of about 0.0002[9] is smaller by one order of magnitude than those obtained at $t' = -2t'' = -0.05$ and $-0.10$. This small value is in apparent disagreement with the fact that Hg1201 and Tl2201 take the highest values of $T_c$ near 100 K among single-$CuO_2$-layer cuprates. We naively assume that $T_c$ increases with an increase in $E_{cond}$ basically.

On the other hand our preliminary VMC calculations with $U = 6 \sim 8$ indicated that $E_{cond}$ increases sharply with increasing $U$ when $t' = -2t'' \sim -0.20$. Ref. 18 showed that SC $E_{cond}$ increases very quickly with increasing $U$ when $U \sim 6$, though at $t' = t'' = 0$ with small lattices. There is in fact a clear indication that $U$ may be larger in Hg1201 and Tl2201 than in LSCO, since their apex oxygens are most remote from the planar Cu site among single-layer cuprates,[10] which should lead to a weakened screening of inter-electron interaction on the $CuO_2$ plane and so to a larger $U$.

Therefore, at a fixed set of $t' = -2t'' = -0.25$ appropriate for Hg1201 and Tl2201,

we investigate in this work more closely whether the increase in $U$ enhances $E_{cond}$ and fills the discrepancy between the theory and the experiment. We have examined the cases of $U = 7 \sim 10$. In the cases of $U = 9$ and $10$, $E_{cond}$ was found to have finite positive bulk-limit values and sharply increase with increasing $U$, as expected. The $E_{cond}$ of the experimental size is obtained with $U \sim 9$. This illustrates that $U$ depends on the crystal structure. We also compute $E_{cond}$ at $t' = -2t'' = -0.18$ with $U = 6$ and $8$ to determine the dependences of $E_{cond}$ on both $U$ and $t' = -2t''$. With both values of $U$, we obtain finite positive values of bulk-limit $E_{cond}$. With $U = 8$, we observe that $E_{cond}$ is larger than the experimental $E_{cond}$ of YBCO. This confirms that with fixed $U$, $E_{cond}$ increases with a decrease in $|t'| = 2t''$ in the region of $|t'| = 2t'' > 0.16$.

Assuming the correlation between $U$ and the apex height $h_{apex}$ we argue that it is understandable that $T_c$ grows with an increase in $h_{apex}$ and also in $|t'| = 2t''$.[10]

In our model, electronic sites are located at the lattice points on a square lattice. Each site has nearest-neighbor sites in the $x$- and $y$-directions separated by a unit length. The site number along both axes is given by $L$. Each boundary facing the $x(y)$- direction continues to another boundary on the opposite side under a periodic (antiperiodic) boundary condition. Our 2D Hubbard model is given by

$$H = -t \sum_{<jl>,\sigma} \left( c^\dagger_{j\sigma} c_{l\sigma} + \text{H.c.} \right) - t' \sum_{<<jl>>,\sigma} \left( c^\dagger_{j\sigma} c_{l\sigma} + \text{H.c.} \right)$$
$$-t'' \sum_{<<<jl>>>,\sigma} \left( c^\dagger_{j\sigma} c_{l\sigma} + \text{H.c.} \right) + U \sum_j c^\dagger_{j\uparrow} c_{j\uparrow} c^\dagger_{j\downarrow} c_{j\downarrow}, \quad (1)$$

where $c^\dagger_{j\sigma}$ ($c_{j\sigma}$) is the creation (annihilation) operator of an electron at site $j$ on the square lattice with a spin $\sigma$; $<jl>$, $<<jl>>$, and $<<<jl>>>$ denote the nearest-, second-, and third-neighbor pairs, respectively; H.c. stands for Hermite conjugate; and $t$, $t'$, $t''$, and $U$ have been defined earlier.

The method of computation is the VMC method with the Jastrow-type trial function;[9] we call it Gutzwiller-Jastrow VMC. Our trial function for the $d$-wave SC state is

$$\Psi_s = \prod_{<jl>} h^{n_j n_l} \cdot P_{N_e} \cdot \prod_i \left[1-(1-g)n_{i\uparrow}n_{i\downarrow}\right] \cdot \prod_k \left(u_k + v_k c^\dagger_{k\uparrow} c^\dagger_{-k\downarrow}\right)|0\rangle, \quad (2)$$

where the fourth factor operates to the vacuum state $|0\rangle$ to generate a BCS-type wave function, in which $c^\dagger_{k\sigma}$ is the Fourier transform of $c^\dagger_{j\sigma}$. The third factor is the Gutzwiller projection operator in which $g$ is the Gutzwiller variational parameter and $i$ labels a site in real space; $n_{i\sigma}=c^\dagger_{i\sigma}c_{i\sigma}$. The second factor $P_{N_e}$ is a projection operator that extracts the component with a fixed total electron number $N_e$; $h$ in the first factor is another variational parameter that optimizes the electronic correlation between n. n. sites. If we set $h = 1$, the trial function is reduced to the Gutzwiller-type. This $h$ factor is substantially effective next to $g$, increasing $E_{\text{cond}}$ by 30-70%.[19] Similar optimizations, however, with respect to the second- and third-neighbor correlations are not so effective so that we do not take them into account. The coefficients $u_k$ and $v_k$ appear in our calculation only in the ratio

$$u_\mathbf{k}/v_\mathbf{k} = \Delta_\mathbf{k} \Big/ \left(\xi_\mathbf{k} + \sqrt{\xi_\mathbf{k}^2 + \Delta_\mathbf{k}^2}\right), \quad (3)$$

where $\Delta_{\bm{k}} = \Delta(\cos k_x - \cos k_y)$ is a $\bm{k}$-dependent $d$-wave gap function with a gap parameter $\Delta$; $\xi_{\bm{k}}$ is defined by

$$\xi_{\bm{k}} = -2t(\cos k_x + \cos k_y) - 4t'\cos k_x \cos k_y - 2t''(\cos 2k_x + \cos 2k_y) - \mu, \qquad (4)$$

where $\mu$ is a variational parameter like chemical potential.

The SC ground state energy is given by

$$E_g = \langle H \rangle \equiv \langle \Psi_s | H | \Psi_s \rangle / \langle \Psi_s | \Psi_s \rangle. \qquad (5)$$

We evaluate this by the Monte Carlo method. We minimize $E_g$, optimizing the variational parameters $g$, $h$, $\Delta$, and $\mu$ by the correlated measurements technique.

The condensation energy $E_{\text{cond}}$ for the SC state is gived as the decrease per site in this $E_g$ for the SC state, $E_g^{(SC)}$, from $E_g^{(normal)}$, the $E_g$ value for the normal state as

$$E_{\text{cond}} = [E_g^{(normal)} - E_g^{(SC)}]/N_{\text{site}}, \qquad (6)$$

where $N_{\text{site}}$ is the number of electronic sites on the lattice. $E_g^{(normal)}$ is obtained variationally using a normal-state trial function that is the trial function in eq. (2) in which the BCS wave function is replaced with the function of the Fermi sea. In this work, for this value, we choose $E_g^{(SC)}(\Delta\sim 0)$ obtained from the Jastrow-type trial function with an infinitesimal value of $\Delta$ and the other optimized parameters that reduce the trial function to the normal state wave function. This way lessens the size effect.

At $t' = -2t'' = -0.25$, we have computed the $E_{cond}$ values in the cases of $U = 7 \sim 10$ and examined the $U$ dependence of $E_{cond}$. We treated lattices with $L = 10 \sim 24$. The results are plotted as a function of $1/L^2$ in Fig. 1. There are two peaks and two dips deviating from the fitting line in Fig. 1; they appear because $\mu$ lies where the density of states is high and low, respectively. The $k$ points are given by $(2\nu, 2\nu'+1)\pi/L$ with $\nu$ and $\nu'$ being integers. In the case where several $k$-points whose $\xi_k$ values are close to $\mu$ align near the Fermi surface in the $k$-space, the density of states near $\mu$ is enhanced, which results in a peak in the plot of $E_{cond}$ vs $1/L^2$. When there is no close $k$ points near the Fermi surface, the opposite occurs, and a dip arises. These are size effects.[20]

In spite of the above-mentioned scattering, the linear fit of $E_{cond}$ as a function of $1/L^2$ is fairly good in the range of $L \geq 10$ so that it allows us to extrapolate the least-square fitting line to $1/L^2 = 0$ corresponding to $L = \infty$. When $U \geq 9$, the extrapolated lines cut the vertical axis at finite positive values. We regard this value as the bulk-limit value of $E_{cond}$. For $U = 8, 9$, and 10 in the limit of $L \to \infty$, we obtained $E_{cond} = -0.0000350$, 0.000332, and 0.00138, respectively. These values are plotted as a function of $U$ in Fig. 2. The bulk-limit $E_{cond}$ is found to increase with increasing $U$ very rapidly. Conversely, the bulk-limit $E_{cond}$ tends to vanish at about $U = 8.0$. The bulk-limit $E_{cond}$ at $U = 6 \sim 8$ are regarded as lying within the error bar at approximately zero. Apparently, there is a threshold value $U_{th}$ of $U$ for the bulk-limit $E_{cond}$ to be finite and positive. $U_{th}$ seems to lie at about 8.0. The $E_{cond}$ at $U = 9$ is near the experimental values of Hg1201 and Tl2201, as discussed later.

We extend our computation to $t' = -2t'' = -0.18$ with $U = 6$ and 8 in order to investigate the dependence of $U_{th}$ on $|t'| (= 2t'')$. This parameter set is close to but outside the medium-$|t'|$ range of $-0.16 \le t' = -2t'' \le -0.08$, where SDW dominates.[9)] In this range, the SDW condensation energy was computed by the VMC method employing the Jastrow-type trial function for the SDW state. In Fig. 1 in ref. 9, the SDW condensation energy with $U = 6$ is plotted against $|t'| = 2t''$ and shown to grow overwhelmingly large in the above-mentioned range. Although the SDW $E_{cond}$ was found to be slightly larger with $U = 8$, the above-mentioned medium-$|t'|$ range where SDW dominates was actually the same in terms of the precision of the mesh of the figure.

As shown in Fig. 3, even with $U = 6$ and 8 the bulk-limit $E_{cond}$ is already finite with 0.000011 and 0.00135, respectively, which are plotted in Fig. 4. With $U = 8$, $E_{cond}$ is already larger that the experimental $E_{cond} \approx 0.0007$ of YBCO. With $U = 6$, the bulk-limit $E_{cond}$ is regarded as zero, lying within the error bar. Therefore, at $t' = -2t'' = -0.18$, we see $U_{th} \approx 6.0$. Data points in Fig. 4 are few but the closeness of $U_{th}$ to 6.0 seems to have a reason since the fitting line in Fig. 3 changes systematically and that for $U = 6$ aims at the coordinate origin. $U_{th}$ increases from 6.0 to 8.0 with a change in $t' = -2t''$ from $-0.18$ to $-0.25$.

Another point is that with a fixed $U$, *e. g.*, at 8, the calculated $E_{cond}$ is found to sharply decrease with an increase in $|t'| = 2t''$, when $|t'| = 2t'' > 0.16$. This is presumably due to deterioration of the Fermi surface nesting.

Incidentally, with $U = 6$ at $t'$ ($= -2t''$) $= -0.05$ and $-0.10$, we obtained finite $E_{cond}$ values of 0.0014 and 0.0010, respectively, in the previous work[9)] so that $U_{th}$ at these band parameter sets are smaller than 6. At $t' = t'' = 0.0$ with $U = 6$, $E_{cond}$ lay within the error bar at zero.

Experimental data of the specific heat $C$ of Tl2201 are available,[21] which allow us to obtain the SC $E_{cond} \cong 0.108$ meV/Cu $\cong 0.000308$ $t$/Cu with $t$ assumed to be 0.35 meV. The peak of $C/T$ is located at ~78 K, significantly lower than $T_c = 93$ K, which allows us to suspect that the $T_c$ of the sample is lower so that we regard the obtained $E_{cond}$ as a lower bound for $E_{cond}$ of the highest-$T_c$ single-layer cuprates. An upper bound for $E_{cond}$ of these cuprates can be obtained from the data of the critical field $H_c(0) = 9300$ Oe for the Hg1223 sample with $T_c = 133.5$ K,[22] which results in $E_{cond} \cong 0.167$ meV/Cu $\cong$ 0.000478 $t$/Cu. $T_c$ is higher than $T_c$ of single-layer cuprates by about 40% and there must be an interlayer interaction. However, since ideal data are lacking, we regard 0.000478 $t$/Cu as an upper bound which we search for $t' = -2t'' = -0.25$. Putting $E_{cond} = 0.000308$ and 0.000478 $t$/Cu in Fig. 2, we obtain approximate estimates of the corresponding $U \cong$ 9.0 and 9.2, respectively for the upper and lower $E_{cond}$ values of Hg1201 and Tl 2201. These values are significantly larger than $U = 7.3$ given by Coldea et al. for $La_2CuO_4$,[6] clearly indicating that the $U$ values of Hg1201 and Tl2201 are larger than that of LSCO.

We have noticed the possibility of having such a large $U \sim 9$ for Hg1201 and Tl2201 from the observation that the height $h_{apex}$ of the apex oxygen from the $CuO_2$ plane is largest among those of single-$CuO_2$-layer cuprates.[10] The latter should weaken the screening of inter-electron interaction due to the remoteness of the apex oxygen anion. The weakened screening should lead to a rise of the effective inter-electron Coulomb interaction $U$ in the $CuO_2$ plane. The decrease in $U$ due to screening by surroundings is very similar to that caused by the relaxation energy of a molecule upon the removal or addition of an electron.[23] Typical magnitude of this energy is 0.4 eV. Since the planar Cu has -2 charged apex oxygen on both sides, it is conceivable that $U$ changes by an order of 1 eV ~ $3t$ depending on $h_{apex}$.

The comparison of $E_{cond}$ between $t'$ ($= -2t''$) $= -0.25$ and $-0.18$ indicates that the increase in $|t'| = 2t''$ decreases $E_{cond}$ and so $T_c$ as well. Apparently, this is in a conflict with the experimental observation that the $T_c$ of single-layer cuprates becomes higher when $|t'| = 2t''$ increases.[10] This conflict can be resolved if we take account of the above-mentioned accompanying effect, *i.e.*, the increase in $U$. It is known from the band theory[10] that the increase in $|t'| = 2t''$ is brought about by the same increase in $h_{apex}$. If the increase in $E_{cond}$ due to the increase in $U$ is rapid enough, it should overcompensate for the negative effect due to the increase in $|t'| = 2t''$. This situation gives rise to positive correlations among $E_{cond}$, $U$, and $|t'|$ and consequently to the experimental systematics between $T_c$ and $|t'|$.

Shih et al. investigated the effect of $t'$ ($= -2t''$) on $T_c$ using the $t$-$J$ model with $J = 0.3$ which means assuming a strong-coupling $U$ fixed at $U \approx 4t^2/J \approx 13.3$.[24] They interpret the SC-pair correlation increasing as a function of $|t'| = 2t''$ up to $\sim 0.30$ as indicating $T_c$ increasing with increasing $|t'| = 2t''$. This behavior suggests a direct correlation of $E_{cond}$ and $|t'|$ which is opposite to the tendency of our $E_{cond}$ which decreases with increasing $|t'| = 2t''$ when $|t'| = 2t'' > 0.16$. There is another difference in that, in their scheme, $T_c$ and $E_{cond}$ should decrease with an increase in $U$ since the corresponding $J \simeq 4t^2/U$ decreases. We naively assume positive correlations among $T_c$, $E_{cond}$, and SC pair functions even in the $t$-$J$ model with $J \sim 0.3$. This is opposite to the $U$ dependence in our scheme. Remarkably, these differences between the two theories are considered to come from the difference in the $U$ regimes that both theories assume. Such difference in behaviors depending on $U$ are qualitatively shown in Fig. 3 in ref. 18 although at $t' = t'' = 0$. An experimental test of two results based on the strong- and intermediate-$U$ couplings is very desirable. Such a test is expected to show which $U$ regime is relevant. It would be very helpful to be able to separate the contributions of the change in $U$, or $J$, and that in $|t'| = 2t''$ by experiments. The outcome of the comparison could disclose a new aspect for searching higher $T_c$ and studying various problems such as the isotope effect.

To explain the high-$T_c$ SC in the framework of the 2D Hubbard model, we have been trying to obtain the SC condensation energy $E_{cond}$ comparable to experimental values for plausible model parameter sets by the Gutzwiller-Jastrow VMC method. In our previous works, we obtained acceptable values of the bulk-limit $E_{cond}$ with $U = 6$ in the small-$|t'|$ region, $e.g.$, at $t'$ $(= -2t'') = -0.05$ and $-0.10$. However, in the large-$|t'|$ region, $e.g.$, at $t' \approx -2t'' \approx -0.25$ this scheme with $U = 6$ did not provide sufficiently large $E_{cond}$ to explain the results of the experiments. In this work, we relaxed the constraint of the fixed $U$ and investigated the $U$ dependence of SC $E_{cond}$, first, at $t' = -2t'' = -0.25$ suitable for Hg1201 and Tl2201. The calculated $E_{cond}$'s were found to take finite positive values in the bulk when $U$ is larger than the threshold $U$ value, $U_{th}$, close to 8. When $U > U_{th}$, $E_{cond}$ increased very rapidly with increasing $U$, and when $U \sim 9.0$ the calculated $E_{cond}$ was estimated to reach the experimental $E_{cond}$ for the two cuprates. Next, we carried out another investigation at $t' = -2t'' = -0.18$. We obtained similar results to those of the above case, with $U_{th} \cong 6.0$ and a rapid increase in $E_{cond}$ with $U > U_{th}$. $E_{cond}$ with a fixed $U$, $e.g.$, at $U = 8$ indicated that $E_{cond}$ decreases with an increase in $|t'| = 2t''$ in the large-$|t'|$ region. We argued that the above-mentioned substantially increased $U \sim 9.0$ is plausibly the reality, induced by the remarkable increase in the apex oxygen height $h_{apex}$ and leading to the agreement of $E_{cond}$ between the theory and the experiment on Hg1201 and Tl2201. This assumption that $U$ changes depending on the crystal structure was pointed out to bring about the experimental observations of the correlations among $T_c$, $h_{apex}$, and $|t'|$.

The authors are grateful to Drs. I. Hase, H. Odagiri, and Y. Tanaka for providing constructive information. Part of this work was carried out using the supercomputing system of KEK. We are grateful to KEK for their support.

1) P. W. Anderson: *The Theory of Superconductivity in High-$T_c$ Cuprate Superconductors* (Princeton Univ. Press, NJ, 1997) Chaps. 1~7.

2) E. Dagotto: Rev. Mod. Phys. **66** (1994) 763.

3) T. Moriya and K. Ueda: Adv. Phys. **49** (2000) 555.

4) J. Kondo: J. Phys. Soc. Jpn. **70** (2001) 808.

5) A. J. Leggett: Nat. Phys. **2** (2006) 134.

6) R. Coldea, S. M. Hayden, G. Aeppli, T. G. Perring, C. D. Frost, T. E. Mason, S. –W. Cheong, and Z. Fisk: Phys. Rev. Lett. **86** (2001) 5377.

7) T. Nakanishi, K. Yamaji, and T. Yanagisawa: J. Phys. Soc. Jpn. **66** (1997) 294.

8) K. Yamaji, T. Yanagisawa, T. Nakanishi, and S. Koike, Physica C **304** (1998) 225.

9) K. Yamaji, T. Yanagisawa, M. Miyazaki, and R. Kadono: Physica C **468** (2008) 1125.

10) A. Pavarini, I. Dasgupta, T. Saha-Dasgupta, O. Jepsen, and O. K. Andersen: Phys. Rev. Lett. **87** (2001) 047003.

11) J. W. Loram, K. A. Mizra, J. R. Cooper, and W. Y. Liang: Phys. Rev. Lett. **71** (1993) 1740.

12) Z. Hao, J. R. Clem, M. W. McElfresh, L. Civale, A. P. Malozemoff, and F. Holtzberg: Phys. Rev. B **43** (1991) 2844.

13) P. W. Anderson: Science **279** (1998) 1196.

14) T. Tohyama and S. Maekawa: Supercond. Sci. Technol. **13** (2000) R17.

15) H. Eisaki, N. Kaneko, D. L. Feng, A. Damascelli, P. K. Mang, K. M. Shen, Z. –X. Shen, and M Greven: Phys. Rev. B **69** (2004) 064512.

16) W. S. Lee, T. Yoshida, W. Meevasana, K. M. Shen, D. H. Lu, W. L. Yang, X. J. Zhou, X. Zhao, G. Yu, Y. Cho, M. Greven, Z. Hussain, and Z.-X. Shen: arXiv: cond-mat/0606347 (2006).

17) N. E. Hussey, M. Abdel-Jawad, A. Carrington, A. P. Mackenzie, and L. Balicas: Nature **425** (2003) 814.

18) H. Yokoyama, Y. Tanaka, M. Ogata, and H. Tsuchiura: J. Phys. Soc. Jpn. **73** (2004) 1119.

19) K. Yamaji, T. Yanagisawa, and S. Koike: J. Phys. Chem. Solids **62** (2001) 237.

20) For each size of the lattice with $L$, we chose an even electron number $N_e$ which makes electron density $\rho = N_e/L^2$ closest to 0.84. The deviation $|\rho-0.84|$ has a tendency to decrease in proportion to $1/L^2$. The average $\langle|\rho-0.84|\rangle \sim 0.0020$ may cause an error in $E_{\mathrm{cond}}$ by 0.00005 at most, according to the $\rho$ dependence of $\rho$.[8] When $L \geq 20$, the error is much smaller. This is well within the error bar due to the VMC computation.

21) J. W. Loram, K. A. Mizra, J. W. Wade, J. R. Cooper, and W. Y. Liang: Physica C **235**-**240** (1994) 134.

22) Y. C. Kim, J. R. Thompson, J. G. Ossandon, D. K. Christen, and M. Paranthaman: Phys. Rev. B 51 (1995) 11767.

23) T. Ishiguro, K. Yamaji, and G. Saito: *Organic Superconductors* (Springer Series in Solid-State Sciences 88, Berlin, Germany, 1998) 2nd ed., Section 8.3.3.

24) C. T. Shih, T. K. Lee, R. Eder, C.–Y. Mou, and Y. C. Chen: Phys. Rev. Lett. **92** (2004) 227002.

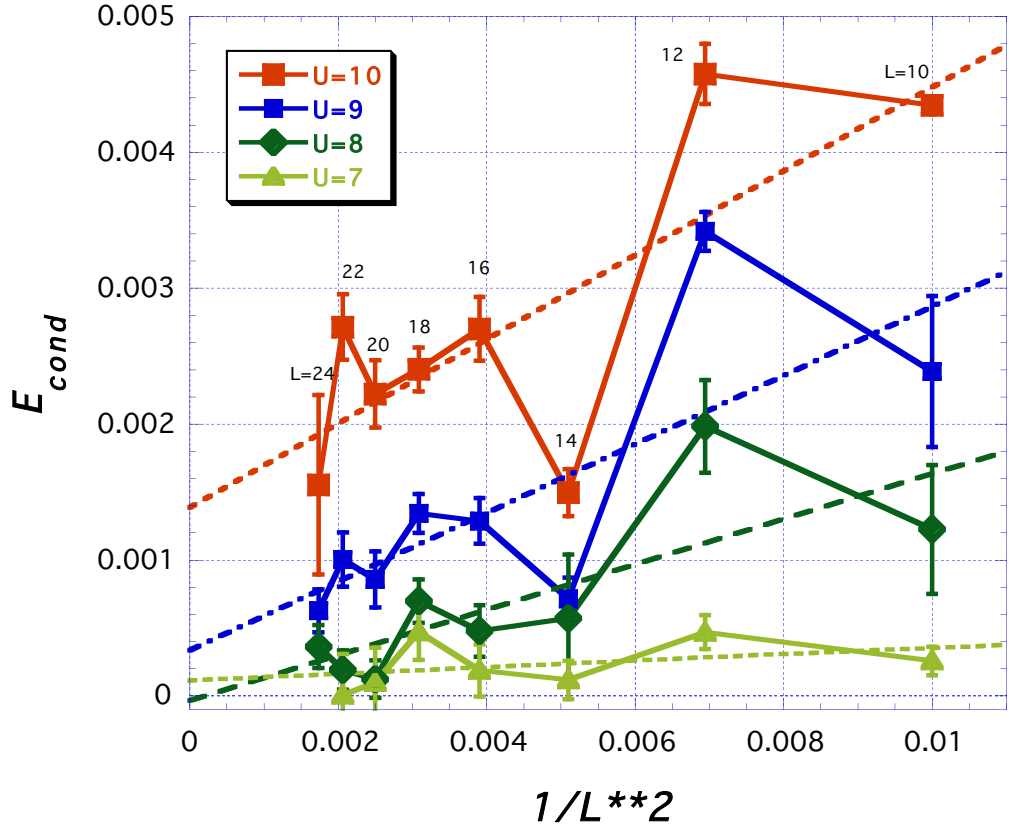

Fig. 1. The condensation energy $E_{cond}$ computed by using the Gutzwiller-Jastrow VMC method is plotted in the case of $t' = -2t'' = -0.25$ with $U = 7 \sim 10$ against $1/L^2$. $L$ is the edge length of the square lattice. The error bar shows the standard deviation. The straight line is the linear fit, which leads to the bulk-limit value in the limit of $L \rightarrow \infty$ as shown by its intercept with the vertical axis.

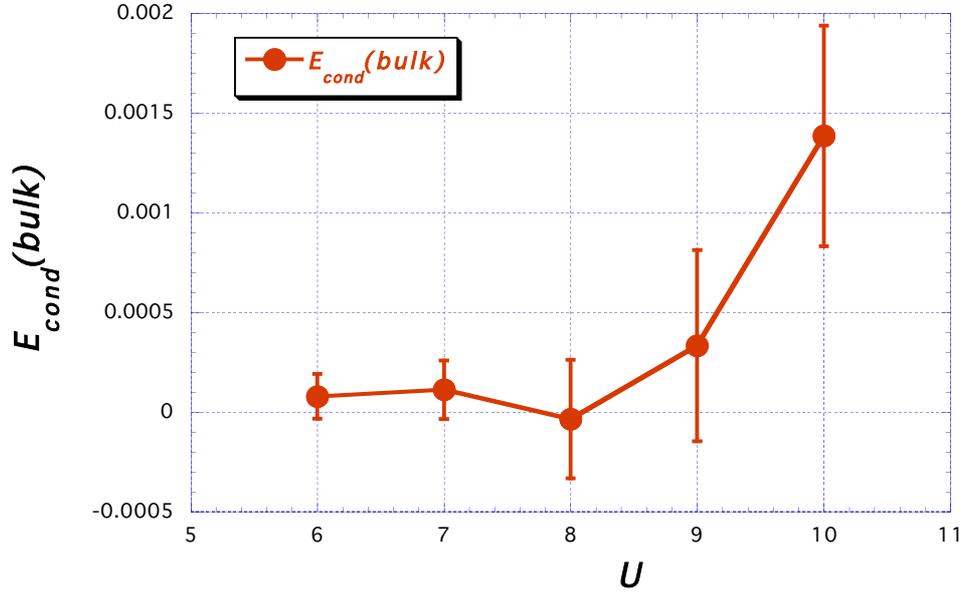

Fig. 2. Bulk-limit $E_{cond}$ is plotted against $U$ in the case of $t' = -2t'' = -0.25$. The error bar shows the standard parameter error in fitting. The $E_{cond}$ at $U = 6 \sim 8$ is regarded to lie within the error bar at approximately zero. Result at $U = 6$ obtained using $E_g^{(normal)}$ is added.

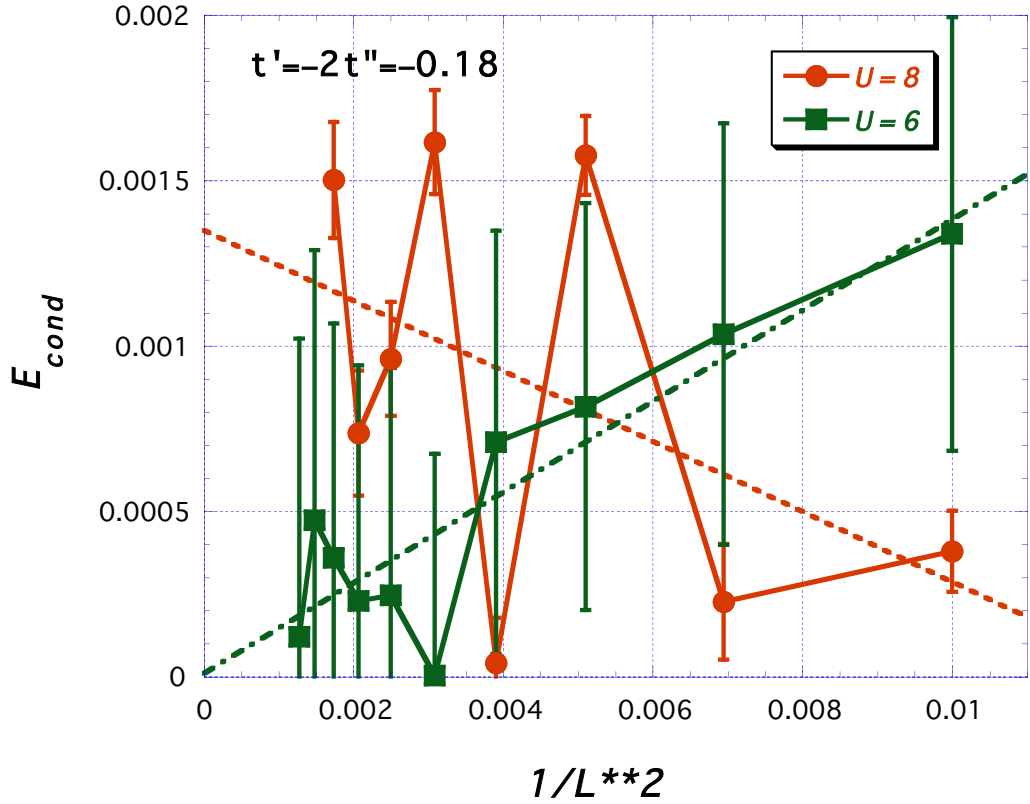

Fig. 3. The condensation energy $E_{cond}$ is plotted in the case of $t' = -2t'' = -0.18$ with $U = 6 \sim 8$ against $1/L^2$. The straight line is the linear fit, which leads to the bulk-limit value in the limit of $L \to \infty$ as shown by its intercept with the vertical axis.

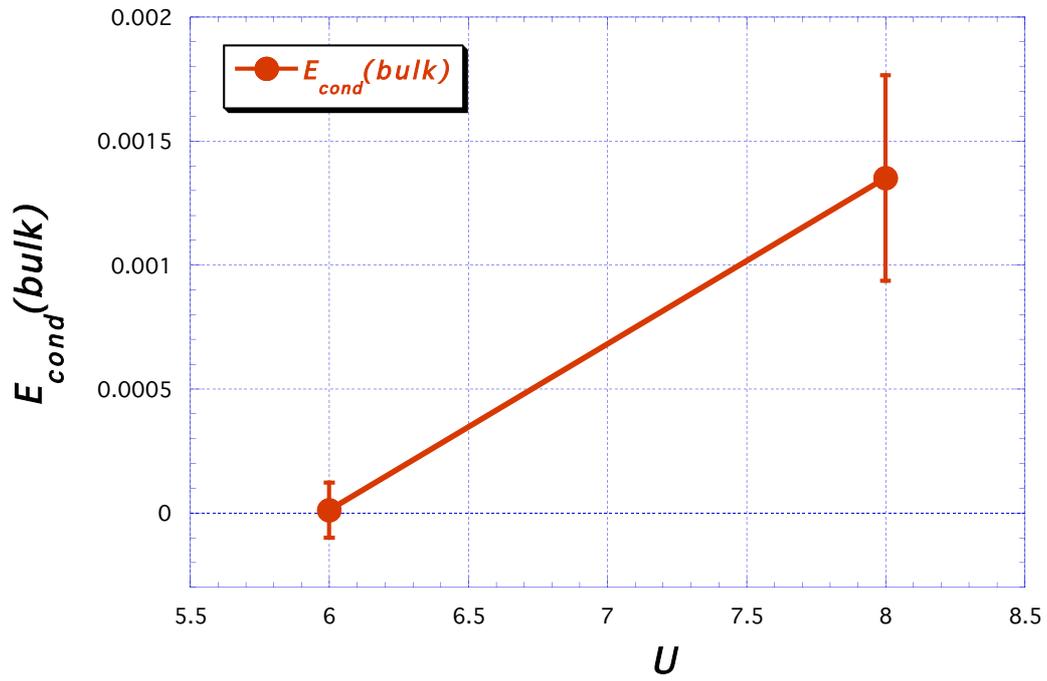

Fig. 4. Bulk-limit $E_{cond}$ is plotted against $U$ in the case of $t' = -2t'' = -0.18$. The fitting line is seen to cross the horizontal axis at approximately $U = 6$.